\documentclass[a4paper,fleqn,usenatbib,useAMS]{mnras}

\usepackage{graphicx}	
\usepackage{amsmath}	
\usepackage{amssymb}	
\usepackage{multicol}        
\usepackage{bm}		
\usepackage{pdflscape}	
\usepackage{multirow}
\let\oldAA\AA
\renewcommand{\AA}{\text{\normalfont\oldAA}}

\usepackage[T1]{fontenc}
\usepackage{ae,aecompl}

\usepackage{color}
\usepackage{amsmath}

\title[GZ: Offset discs and bars in SDSS]{Galaxy Zoo: Finding offset discs and bars in SDSS galaxies}

\author[Sandor Kruk et al.]{
Sandor J. Kruk,$^{1}$\thanks{E-mail: \href{mailto:sandor.kruk@physics.ox.ac.uk}{sandor.kruk@physics.ox.ac.uk}}
Chris J. Lintott,$^{1}$ 
Brooke D. Simmons,$^{1,2}$\thanks{Einstein Fellow}
Steven P. Bamford,$^{3}$ 
\newauthor
Carolin N. Cardamone,$^{4}$ 
Lucy Fortson,$^{5}$ 
Ross E. Hart,$^{3}$ 
Boris H{\"a}u{\ss}ler,$^{6}$
\newauthor 
Karen L. Masters,$^{7}$
Robert C. Nichol,$^{7}$ 
Kevin Schawinski,$^{8}$
Rebecca J. Smethurst$^{1,3}$
\thanks{This investigation has been made possible by the participation of over 350,000 users in the Galaxy Zoo project. Their contributions are acknowledged at \href{http://authors.galaxyzoo.org}{http://authors.galaxyzoo.org}}\\
$^{1}$Oxford Astrophysics, Department of Physics, University of Oxford, Denys Wilkinson Building, Keble Road, Oxford, OX1 3RH, UK\\
$^{2}$Center for Astrophysics and Space Sciences (CASS), Department of Physics, University of California, San Diego, CA 92093, USA\\
$^{3}$School of Physics and Astronomy, The University of Nottingham, University Park, Nottingham NG7 2RD, UK\\
$^{4}$Math and Science Department, Wheelock College, 200 The Riverway, Boston, MA 02215, USA\\
$^{5}$School of Physics and Astronomy, University of Minnesota, 116 Church St. SE, Minneapolis, MN 55455, USA\\
$^{6}$ESO - European Southern Observatory, Alonso de Cordova 3107, Vitacura, Santiago, Chile\\
$^{7}$Institute of Cosmology and Gravitation, University of Portsmouth, Dennis Sciama Building, Barnaby Road, Portsmouth, PO1 3FX, UK\\
$^{8}$Institute for Astronomy, Department of Physics, ETH Z{\"u}rich, Wolfgang-Pauli Strasse 27, CH-8093 Z{\"u}rich, Switzerland}

\date{Accepted 2017 April 26. Received 2017 April 26; in original form 2017 January 17}

\pubyear{2017}

\begin{document}
\label{firstpage}
\pagerange{\pageref{firstpage}--\pageref{lastpage}}
\maketitle

\begin{abstract}
We use multi-wavelength SDSS images and Galaxy Zoo morphologies to identify a sample of $\sim$$270$ late-type galaxies with an off-centre bar. We measure offsets in the range 0.2-2.5 kpc between the photometric centres of the stellar disc and stellar bar. The measured offsets correlate with global asymmetries of the galaxies, with those with largest offsets showing higher lopsidedness. These findings are in good agreement with predictions from simulations of dwarf-dwarf tidal interactions producing off-centre bars. We find that the majority of galaxies with off-centre bars are of Magellanic type, with a median mass of $10^{9.6} M_{\odot}$, and 91\% of them having $M_{\star}<3\times10^{10} M_{\odot}$, the characteristic mass at which galaxies start having higher central concentrations attributed to the presence of bulges. We conduct a search for companions to test the hypothesis of tidal interactions, but find that a similar fraction of galaxies with offset bars have companions within 100 kpc as galaxies with centred bars. Although this may be due to the incompleteness of the SDSS spectroscopic survey at the faint end, alternative scenarios that give rise to offset bars such as interactions with dark companions or the effect of lopsided halo potentials should be considered. Future observations are needed to confirm possible low mass companion candidates and to determine the shape of the dark matter halo, in order to find the explanation for the off-centre bars in these galaxies.

\end{abstract}





\section{Introduction}

Bars are common in disc galaxies, between one and two thirds of local disc galaxies being barred \citep{Sellwood1993,Sheth2008,Masters2011}, depending on the bar classification method and the wavelengths in which the galaxies are observed. Some of these galaxies exhibit a peculiar feature, a bar that appears to be offset from the photometric centre of the galaxy discs. 

Such an offset seems common in low mass late-type galaxies of the kind \citet{Vauc1972} defined as Magellanic spirals after their prototype, the Large Magellanic Cloud (LMC) \citep{deVauc1955}. The nearest such galaxy and the best-known example, the LMC itself hosts a bar that is offset from the centre of the outer disc isophotes by $\sim$$0.4$ kpc, while the kinematic centre of the HI disc is offset from both by as much as $\sim$$ 0.8$ kpc \citep{vanMarel2001}.  

The origin of the off-centre bar in the LMC is not well understood. \citet{Zhao2000} suggested that the bar in the LMC is off-centre as a consequence of a recent tidal interaction with the Small Magellanic Cloud (SMC) and the Milky Way. Numerical simulations of barred galaxies have shown that a bar may become offset from the disc following an interaction with a companion, while the disc of the galaxy becomes lopsided \citep{Athanassoula1996, Athanassoula1997,Berentzen2003,Besla2012,Yozin2014}. Recently, \citet{Pardy2016} have followed up on the idea of a tidally induced offset in barred Magellanic type galaxies using N-body and hydrodynamic simulations of dwarf-dwarf galaxy interaction. They investigated the relation between the dynamical, stellar and gaseous disc centres and the bar in a 1:10 mass ratio interaction, characteristic of the interaction between the SMC and the LMC (the stellar mass of the LMC is $3\times10^9 M_{\odot}$ \citep{vanderMarel2002}, while that of the SMC is $3\times10^8 M_{\odot}$ \citep{SMC2004}). They conclude that an offset between the photometric centre of the bar and the photometric centre of the disc is produced in such an interaction. The predicted shift is, at most, 1.5-2.5 kpc depending on the details of the interaction and type of halo considered. The largest offsets are produced for smaller impact parameters for the passing galaxy and large inclination angles with respect to the plane of the primary galaxy. The amplitude of the subsequent offset is correlated with the distorted asymmetry (lopsidedness) of the disc and it decreases with time, with the distortions vanishing after 2 Gyr. Surprisingly, they find that the stellar bar is always coincident with the dynamical centre and it is the disc that is displaced from the dynamical centre (see, e.g., Figure 3 in \citealt{Pardy2016}).

Offset bars are observed in other galaxies in the local Universe as well. In a first large scale study of such nearby galaxies, \citet{Feitzinger1980} measured an average offset between the centre of the bar and the disc of $0.8$ kpc for 18 galaxies. More recently, \citet{Swardt2015} measured an offset of $\sim$$0.9$ kpc between the centre of the stellar bar and the centre of the disc in NGC3906. In this case the bar centre coincides with the dynamical centre determined through HI observations. In contrast with the LMC, NGC3906 is observed to be isolated, thus a possible explanation for the observed offset is an interaction with the dark matter subhalo, or an unidentified fast moving companion.  \citet{Bekki2009} suggested that dark satellites with the mass of $10^{8}-10^{9} M_{\odot}$ and either no or very little observable matter can create an offset bar in a collision with a Magellanic type galaxy. Alternatively, modeling of lopsided galaxies suggests that long-lived off-centre bars and asymmetries may be a consequence of misalignments between the stellar disc and halo \citep{Jog1997,levine1998,Noordermeer2001}. The lopsidedness in the stellar disc can be caused by several phenomena, such as tidal interactions \citep{Bale1969}, gas accretion (\citet{Zaritsky1997, Bournaud2005}) or small asymmetries in the galactic halo \citep{Jog2009}.

Despite the availability of large surveys, observationally the origin of offsets and the asymmetries in Magellanic type galaxies has not yet been established. There has been contradictory evidence about the frequency of the companions of this type of galaxy. In a large survey of local Magellanic type galaxies, \citet{Odewahn1994} found that 71 out of 75 galaxies have a nearby neighbour, within a projected separation of 120 kpc. In contrast, in an HI follow-up study of a subset of the Magellanic type barred galaxies observed by \citet{Odewahn1994}, \citet{Wilcots2004} found that only 2 of 13  were interacting with their neighbour, clearly affecting their morphologies.

In this paper, we conduct the first systematic search for galaxies with offsets between the stellar bar and the discs in the largest survey in the local Universe, the Sloan Digital Sky Survey (SDSS) \citep{York2000}. With visual classifications from the Galaxy Zoo citizen science project \citep{Lintott2008, Willett2013}, we are able to identify a large sample of local barred galaxies. Using 2D parametric decomposition we can decompose the galaxies into individual components (bars, discs and bulges), measure the offsets between the bars and the discs and quantify the disc asymmetry. Therefore, we are able to identify a sample of galaxies with offset bars and study their individual properties, as well as search for companions to identify the cause of the offsets. Throughout the paper we adopt the WMAP Seven-Year Cosmological parameters \citep{Jarosik2011} with ($\Omega_{M},\Omega_{\Lambda},h) = (0.27,0.73,0.71)$.

\section{Data}

\begin{figure*}
 \includegraphics[width=0.7\textwidth]{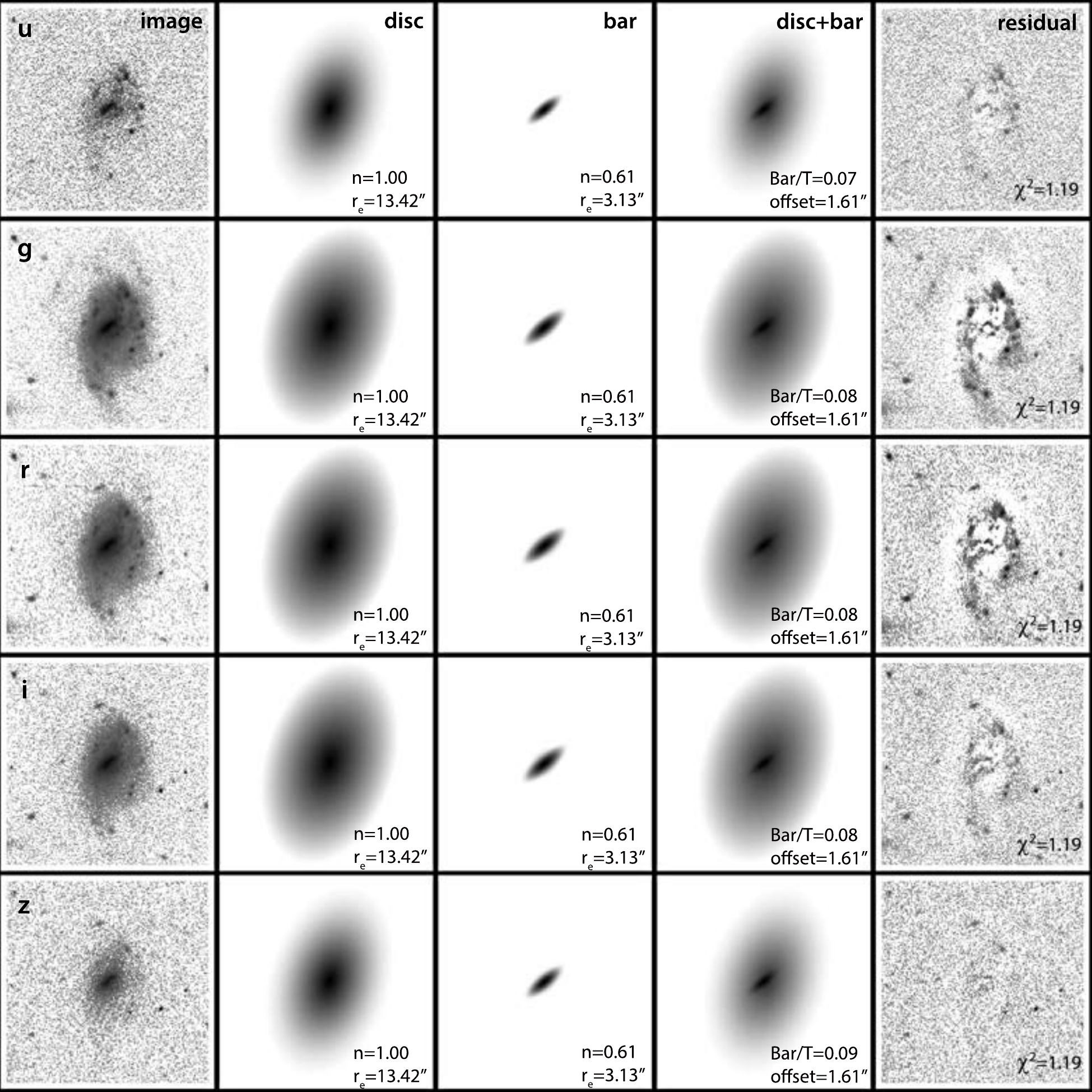}
 \caption{Images of galaxy J143758.75+412033.0 in \textit{u, g, r, i, z} bands. Example of a GALFITM disc+bar fit, model and residuals in the 5 bands, used to identify galaxies with offset bars. The first column shows the original images, the second shows the model for the exponential disc and the third column the bar as a free S\'ersic component. The fourth column is the combined bar+disc model and the last column shows the residual and the reduced $\chi^2$. The projected offsets were measured as the separation on the sky between the centres of the two components. The legend shows the S\'ersic index, the effective radius for each component and the bar-to-total luminosity ratio in the five bands.}
 \label{example}
\end{figure*}

All the galaxies used in the study are drawn from the Sloan Digital Sky Survey (SDSS) DR7 \citep{SDSSDR7, Strauss2002}. We use visual classifications of galaxy morphologies from the Galaxy Zoo 2\footnote{\url{http://zoo2.galaxyzoo.org}} project (GZ2) \citep{Willett2013} which asked citizen scientists to provide detailed information about the visual appearance of galaxies. The full question tree for each galaxy image is shown in Figure 1 of \citet{Willett2013}. 

From the superset of 240,419 galaxies classified in GZ2\footnote{Data available from \url{http://data.galaxyzoo.org}} and with stellar masses available from the MPA-JHU catalogue \citep{Kauffmann2003a}, we have selected all the galaxies with spectroscopic redshifts 0.005<$z$<0.06, a redshift range with reliable GZ2 morphological classifications and suitable SDSS image resolution. Identifying bars in highly inclined galaxies is challenging, thus we selected only galaxies with an axis ratio of $b/a>0.5$ given by the exponential model fits in SDSS \citep{Stoughton2002}, corresponding to inclinations $i\lesssim60^{\circ}$.

In order to reach the bar question a Galaxy Zoo user must first classify a galaxy as a non edge-on galaxy with a disc or features. Following \citet{Masters2011}, we only selected galaxies for which there were at least 10 answers to the question `Is there a sign of a bar feature through the centre of the galaxy?'. Throughout this paper we will be using the debiased likelihoods, denoted as $p_{bar}$, from \citet{Willett2013}. A galaxy was classified as being barred if the number of volunteers identifying it as having a bar is larger than, or equal to the number identifying it as not having a bar, i.e. $p_{bar}\geq0.5$. The selection resulted in a large sample of 5,485 barred galaxies. 

The selection of barred galaxies with $p_{bar}\geq0.5$ has been shown to pick up predominantly intermediate and strong bars, when compared to expert classifications such as those in \citet{Nair2010}, as discussed in Appendix A of \citet{Masters2012} and also shown in Figure 10 in \citet{Willett2013}. This cutoff was chosen as an unavoidable compromise between having a sample with high purity and a complete sample of barred galaxies. Lowering the cutoff would increase the completeness of the sample by including a higher fraction of weak bars, but would also contaminate the sample with non-barred galaxies.

To avoid problems with deblending we exclude merging or overlapping galaxies. According to \citet{Darg2010}, in GZ1 \citep{Lintott2011} this can be done with a cut of the GZ merging parameter $p_{merg} < 0.4$. The galaxies in GZ2 are a subsample of the galaxies classified in GZ1, and although using a different classification tree, $p_{merg}$ has a similarly strong correlation with the projected galaxy separation, as shown by \citet{Casteels2013}. Our final, large sample of barred galaxies contains 5,282 galaxies. Each galaxy was inspected by at least 19 volunteers and the mean number of classifications per galaxy was ~42. We also make use of volunteers' classifications of the galaxy bulges, as described in \citet{Simmons2013}. The volunteers were asked to classify the bulges of these systems into four categories: \textsc{No-bulge}, \textsc{Just-noticeable}, \textsc{Obvious}, \textsc{Dominant}. We split the sample into two categories: `disc dominated' (having combined debiased likelihoods for \textit{no-bulge}+\textit{just-noticeable}>\textit{obvious}+\textit{dominant}) and `obvious bulges' (with \textit{no-bulge}+\textit{just-noticeable}<\textit{obvious}+\textit{dominant}). There are 2,625 `disc dominated' galaxies (50\% of the sample) and 2,657 galaxies with `obvious bulges' (50\% of the sample).

\section{Measuring bars and discs}

\subsection{Galaxy image decomposition}

A key observable is the spatial distribution of light in a galaxy, which can be modeled using parametric functions such as the S\'ersic profile. In a subsequent paper (Kruk et al, in prep.) we will discuss in detail the 2D decomposition method used to fit the full sample of $\sim$$5,000$ barred galaxies with three components using GALFITM\footnote{GALFITM is publicly available at \url{http://www.nottingham.ac.uk/astronomy/megamorph/}}, developed by the MegaMorph project \citep{Bamford2011,Heussler2013}. GALFITM is a modified version of GALFIT3.0 \citep{Peng2010} that makes use of the full wavelength coverage of surveys and enables fitting across multiple wavelengths, in order to increase the accuracy of measured parameters. This is achieved by setting each parameter of the model to be a polynomial function of wavelength. GALFITM then optimises the coefficients of these polynomials to best match the multi-band data. As a result, it improves the effective radius and  S\'ersic index \textit{n} estimates in low-S/N bands and, consequently, it improves the photometry of fainter components. The multi band fitting was applied to bulge-disc decompositions of 163 artificially redshifted nearby galaxies and shown to improve the measurements of structural parameters \citep{Vika2014}. 

In this study, we use publicly available data from SDSS in five bands (\textit{ugriz}). To account for seeing, GALFITM convolves the model with a PSF, provided by SDSS for each passband. To ensure that only the targeted galaxies are fitted, we created a mask for each galaxy field in the \textit{r}-band using \textit{\uppercase{SE}xtractor} \citep{Sextractor} and we used it for all the 5-bands in the fitting process.

The galaxy model included discs, bars and bulges chosen according to the visual classifications from Galaxy Zoo, as detailed in Section 2.  We fitted the galaxies in the `disc dominated' and in the `obvious bulges' samples with two (disc+bar) and three components (disc+bar+bulge), respectively, using an iterative process. First, we fitted a single S\'ersic profile, with the purpose of providing initial values for the parameters for the subsequent fits. Subsequently, we fitted a simple, two component model: an exponential disc and a bar with a free S\'ersic index, using the estimated parameters from the single S\'ersic fit as initial guesses. The bar component was modeled as an ellipse with an initial axis ratio of 0.2, an initial S\'ersic index of 0.7, a smaller effective radius than the disc and dimmer, but initially having the same centre as the disc. The centres of the two components were allowed to vary freely from each other across the image, without constraints, allowing offsets between them. This was the final step in the case of `disc dominated' galaxies, while for the `obvious bulges' sample we added a third component, a bulge. The bulge was modeled as a smaller round component, centred on the bar and with an initial S\'ersic index of $n=2$. We constrained the centres of the bulge and the bar to be the same, to avoid the third component converging to a nearby clump or overlapping star that has not been masked out. This is motivated both by visual inspection and physical processes: bars are thought to channel gas and build up bulges at their centres \citep{Kormendy2004}.

The magnitudes of the components were allowed to vary freely with wavelength, while the S\'ersic indices and the effective radii were kept fixed across the five bands. This effectively means that there are no colour and, hence, no stellar population gradients within the models of the individual components, which is a simplified picture of galaxy structure. Our assumption is justified as a first approximation as we are primarily interested in determining the centres of the components by using all the five bands. 

The two-component fits converged for 2,186 (83\%) and the three-component fits for 2,205 galaxies (83\%). An example of a two-component model (disc+bar) for a galaxy, the images and residuals, in five bands, can be seen in Figure \ref{example}. In some cases a second disc component was fitted instead of a bar, thus we excluded galaxies that had a second component with an axis ratio $b/a>0.6$ (500 galaxies). We also excluded galaxies with discs having unphysically large effective radii, $r_{e}>200$ pixels (170 galaxies), corresponding to 8 kpc at $z=0.005$, 16 kpc at $z=0.01$ and 91 kpc at $z=0.06$. Although 8-16 kpc are plausible values for the disc effective radius, there were only two galaxies discarded between the redshifts $0.005<z<0.01$, both of which showed unrealistically large $r_{e}$'s when inspected. We also excluded bar and bulge components with too large S\'ersic indices, $n>8$ (176 galaxies), as these are unphysical values and do not represent a good model of the bar and bulge. Finally, we excluded 188 galaxies where a clump or a foreground overlapping star was fitted instead of one of the components. The bar was assumed to be the component with lower axis ratio of the two components fitted, which should be the case as the galaxies were selected to be face-on. We inspected all the images and checked if the disc+bar and the disc+bar+bulge models were good representations of the target galaxy, by checking that the bar (and the bulge, in some cases) was not visible in the residuals. The final sample that was successfully fitted consists of 3,357 galaxies (a success rate of 64\%): 1,532 galaxies with disc+bar (`disc dominated sample') and 1,825 galaxies with disc+bar+bulge (`obvious bulge' sample). Henceforth we refer to the subsample of galaxies with meaningfully converged disc+bar and disc+bar+bulge fits collectively as the \textsc{fitted-bar sample}. The galaxies where the fits failed and those that were subsequently excluded have a similar distribution of $p_{bar}$ as the \textsc{fitted-bar sample} (with a maximum of 10\% difference at $p_{bar}\sim0.5$), hence the final sample is not biased by excluding 36\% of the barred galaxies.

\begin{figure}
 \includegraphics[width=\columnwidth]{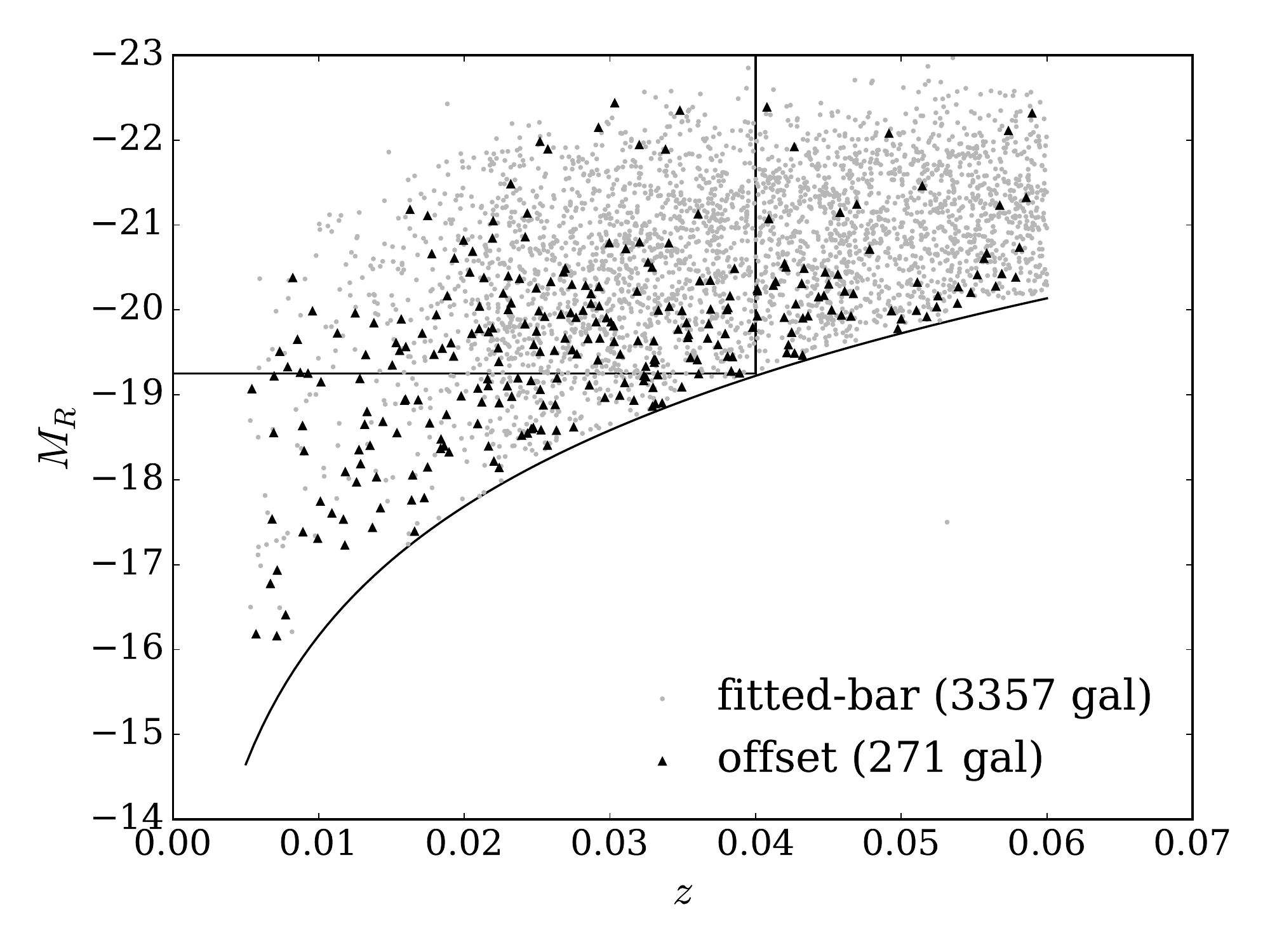}
 \caption{The \textit{r}-band Petrosian absolute magnitudes of the samples used in the paper: the \textsc{fitted-bar sample} and the \textsc{offset sample}, as identified in Section 3.2. The box contains the galaxies in the \textsc{volume-limited sample} (1,583 galaxies) as defined in Section 4.3. The curved line corresponds to the GZ2 completeness limit of 17 magnitudes, at a particular redshift.}
 \label{sample_selection}
\end{figure}

\subsection{Offset Sample}

\begin{figure*}
 \includegraphics[width=\textwidth]{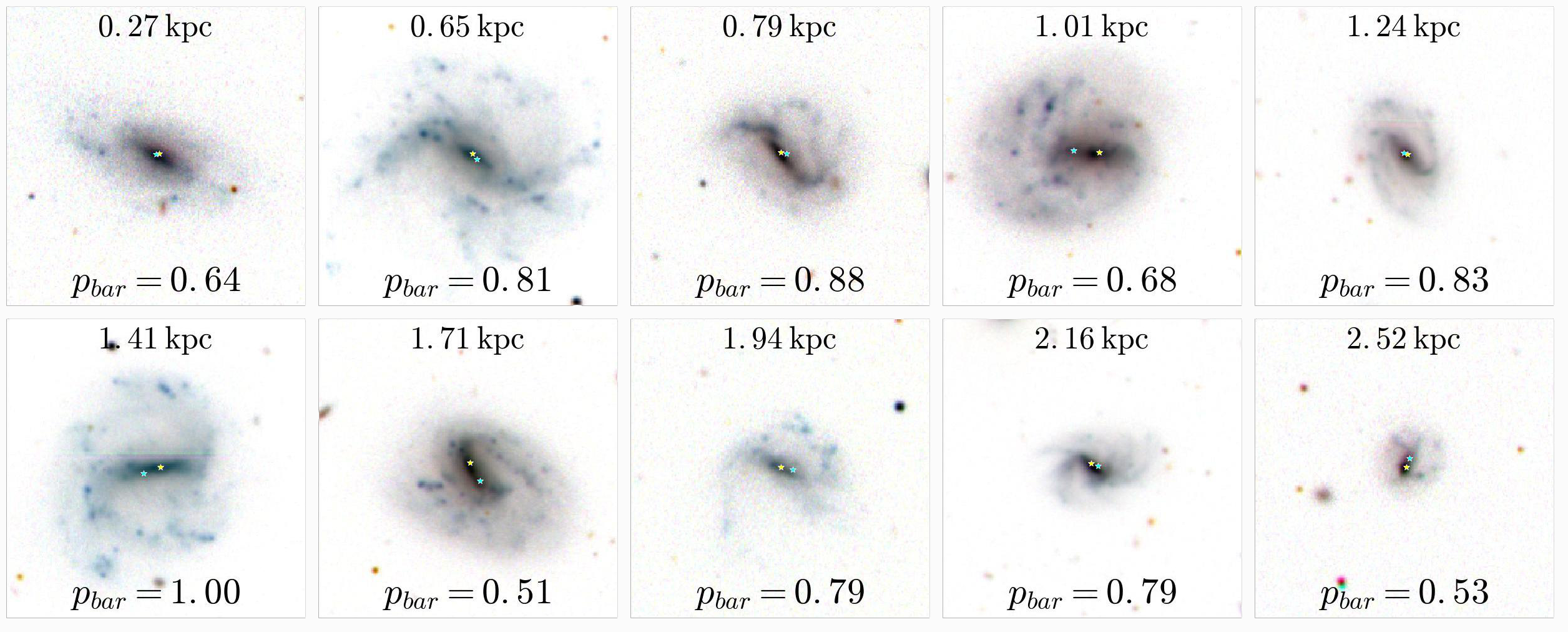}
 \caption{Examples of galaxies with offset discs and bars in SDSS; inverted colour \textit{gri} composite images. The measured deprojected photometric offset between the bar and the disc is given at the top of each image. The GZ2 debiased likelihood that the galaxy has a bar is given at the bottom of each image. The centre of the bar component, according to the best fit model, is marked with a yellow star, while the photometric centre of the disc is marked with a cyan star. The images are 1 arcmin x 1 arcmin.}
 \label{offset_mosaic}
\end{figure*}

We calculated the offsets between the disc and the bar as the projected distance between the photometric centres of the two components. If the measured offset between the photometric centres of bar and disc components is larger than the FWHM of the PSF, we consider the galaxy to have an offset bar. In SDSS, the FWHM of the PSF varies between different fields and bands \citep{Bramich2012}. In the frames used in this study it ranged between $0.83\arcsec-2.33\arcsec$ in the \textit{u}-band (with a median of $1.34\arcsec$) and $0.56\arcsec-1.99\arcsec$ in the \textit{i}-band (with a median of $1.06\arcsec$). Since we fitted five bands simultaneously, we considered a galaxy to have an offset bar if the projected offset was larger than the smallest FWHM of the PSF of the five bands. In the majority of cases, this was the \textit{i}-band or \textit{r}-band. This cut for identifying galaxies with off-centre bars in five bands is conservative since the bar is not a round feature, similar to the shape of the PSF, but rather an extended, linear feature. We are, therefore, identifying the galaxies in our sample with the largest offsets.

The measured offsets were then converted into a physical offset at the position of the galaxy, and deprojected, adopting a simple analytical 1D approximation used to deproject bars \citep{Martin1995, Gadotti2007}. The deprojected offset is:

\begin{equation}
d^{\mathrm{dep}}_{\mathrm{offset}}= d^{\mathrm{proj}}_{\mathrm{offset}} \sqrt{\sin{\alpha}^2 \sec{i}^2+\cos{\alpha}^2}
\label{eqn}
\end{equation}

\noindent where $\alpha$ is the angle between the projected major axis of the bar and the inclined disc (the difference in the position angles of the two components) and $\sec{i} \sim 1/(b/a)_{\mathrm{disc}}$. The uncertainties in the deprojections are small ($\sim$$20\%$), since the galaxies were selected such that $i\lesssim 60^{\circ}$ \citep{Zou2014}, and since the sizes of the offsets are small compared to the sizes of bars and discs.

GALFITM also returns the errors in the estimated parameters for a particular model, which is typically of the order of a few percent in the estimate of the offset. GALFIT errors are known to underestimate the true error because it assumes uncorrelated noise and it does not account for contribution from systematic model errors, as shown by \citet{Haeussler2007}. 

Inspecting the images of galaxies found to be offset, we observed that the majority of them were blue, and therefore, young with a bar and one or more spiral arms, with an offset between the stellar bar and disc being clearly noticeable. We found a sample of 271 galaxies having bars offset from the photometric centre of the disc, most of them faint, as shown in Figure \ref{sample_selection}, in comparison to the \textsc{fitted-bar sample}. 87\% of these galaxies have projected offsets larger than $1\arcsec$, which corresponds to 0.1 kpc at z=0.005 or 1.1 kpc at z=0.06.  Therefore we are able to detect similar offsets to those suggested by \citet{Pardy2016}. Henceforth we refer to this sample of 271 galaxies, as the \textsc{offset sample}. This is currently the largest sample of such galaxies. Some examples of galaxies with offset bars can be seen in Figure \ref{offset_mosaic} and the results from the parametric fitting are summarized in Table \ref{fitting_results}. For comparison, we also select a mass and redshift-matched \textsc{comparison sample} of 271 galaxies with centred bars (selected such that the projected offset is smaller than the PSF FWHM).

\subsection{Quantifying lopsidedness}

In addition to measuring offset distances between the bar and disc components, we also measured the lopsidedness of each galaxy. According to \citet{Peng2010}, this asymmetry can be quantified by expressing the shape of a galaxy as a Fourier perturbation on a perfect ellipse:

\begin{equation}
r(x,y) = r_{0}(x,y) \Big (1+ \sum_{m=1} ^{N}a_{m} \cos(m\phi+\phi_{m}) \Big ),
\end{equation}

\noindent  where $r_{0}(x,y)$ is the radial coordinate of a traditional ellipse,  $\phi_{m}$ denotes the phase of the $\textit{m}$ component and the amplitude of the Fourier component is defined as $A_{m}=|a_{m}|$. The amplitude of the first Fourier mode ($\textit{m}=1$), $A_{1}$ quantifies the lopsidedness of the galaxy disc, the variation in the size of the effective radius on opposing sides of the galaxy. The amplitude of the second Fourier mode ($\textit{m}=2$) $A_{2}$ quantifies the strength of the distortions by structures which have symmetry on rotation by $180^{\circ}$, such as bars or spiral arms. 

To study the lopsidedness of the galaxies, we measured the $A_{1}$ amplitude by fitting an $m_{1}$ Fourier mode on an exponential profile using GALFITM. A high $A_{1}$ amplitude suggests that the photometric centre of an irregular galaxy is not located at the centre of the galaxy, modeled as an ellipse (geometric centre). Therefore, if the mean peak intensity is located in the bar component, galaxies with offset bars should show large $m_{1}$ amplitudes.

\section{Results}

\subsection{Bar-Disc Offsets}

\begin{table*}
\center
\begin{tabular}{lcccccccccccc}
\hline\hline
SDSS Name & Redshift & $m_{r}$ & \multicolumn{2}{c}{Disc} & \multicolumn{2}{c}{Bar}  &  \multicolumn{2}{c}{Bulge} & $\log(M_{\star}) $ & $A_{1} $ & Offset & Offset\\
& &[mag] & \textit{$r_{e}$} [kpc] &  \textit{n} &  \textit{$r_{e}$} [kpc] & \textit{n} &  \textit{$r_{e}$} [kpc] &  \textit{n}  & $[M_{\odot}]$  & & [arcsec] &  [kpc]\\\hline
J001723.39-003112.8 & 0.032 & 16.71 & 3.13  & 1.00  & 1.25  & 0.49 & - & - & 9.40 & 0.20 &  2.80 & 1.98\\
J163037.96+272744.2  & 0.059 & 14.96 & 11.35 & 1.00 & 6.33  & 0.48  & 0.77 & 1.03 & 11.07 & 0.09 & 0.97 & 1.24\\
J023356.29+005525.2  & 0.022 & 15.17 & 5.27 & 1.00  & 1.24  & 0.51 & - & - & 9.59 & 0.08 & 1.19 & 0.58\\
J102003.64+383655.9  & 0.007 & 13.87 & 2.60  & 1.00  & 0.80  & 0.87 & - & - & 9.05 & 0.28 & 6.84 & 1.01\\
J074951.23+184944.3  & 0.016 & 14.78 & 6.11  & 1.00 & 1.22  & 0.25  & - & - & 9.34 & 0.07  & 1.08 & 0.38\\
J132743.83+624559.6  & 0.022 & 13.93 & 8.93  & 1.00 & 4.23  & 0.70  & 0.49 & 1.38 & 10.54  & 0.04 & 1.90 & 1.19\\
J155946.42+371437.9 & 0.057 & 16.74 & 8.56 & 1.00  & 1.46  & 2.56  & - & - & 9.91 & 0.17 & 1.22 & 1.34\\
J111041.31+585646.5 & 0.046 & 16.42 & 5.02  & 1.00 & 3.10  & 0.97  & - & - & 9.93 & 0.18 & 1.23 & 1.19\\
J134308.83+302015.8 & 0.035 & 13.66 & 12.61  & 1.00 & 7.19  & 0.26  & 0.96 & 0.43 & 11.09 & 0.07 & 1.16 & 1.00\\
J165214.37+635738.9 & 0.017 & 14.71 & 3.75  & 1.00 & 0.91  & 0.10  & - & - & 9.77 & 0.18 & 3.25 & 1.22\\
\hline
\end{tabular}
\caption{Properties for 10 out of the 271 galaxies in the \textsc{offset sample}, fitted with disc+bar or disc+bar+bulge components. The redshifts and \textit{r}-band apparent Petrosian magnitudes were drawn from SDSS DR7 and the stellar masses were drawn from the MPA-JHU catalogue \citep{Kauffmann2003a}. The disc component was fitted with an exponential profile ($n=1$), while the bar and bulge with a free S\'ersic profile. The offsets were measured between the photometric centres of the disc and of the bar and the physical offsets were deprojected using Equation \ref{eqn}. Full table is available in the electronic version of the paper.}
\label{fitting_results}
\end{table*}

\begin{figure}
 \includegraphics[width=\columnwidth]{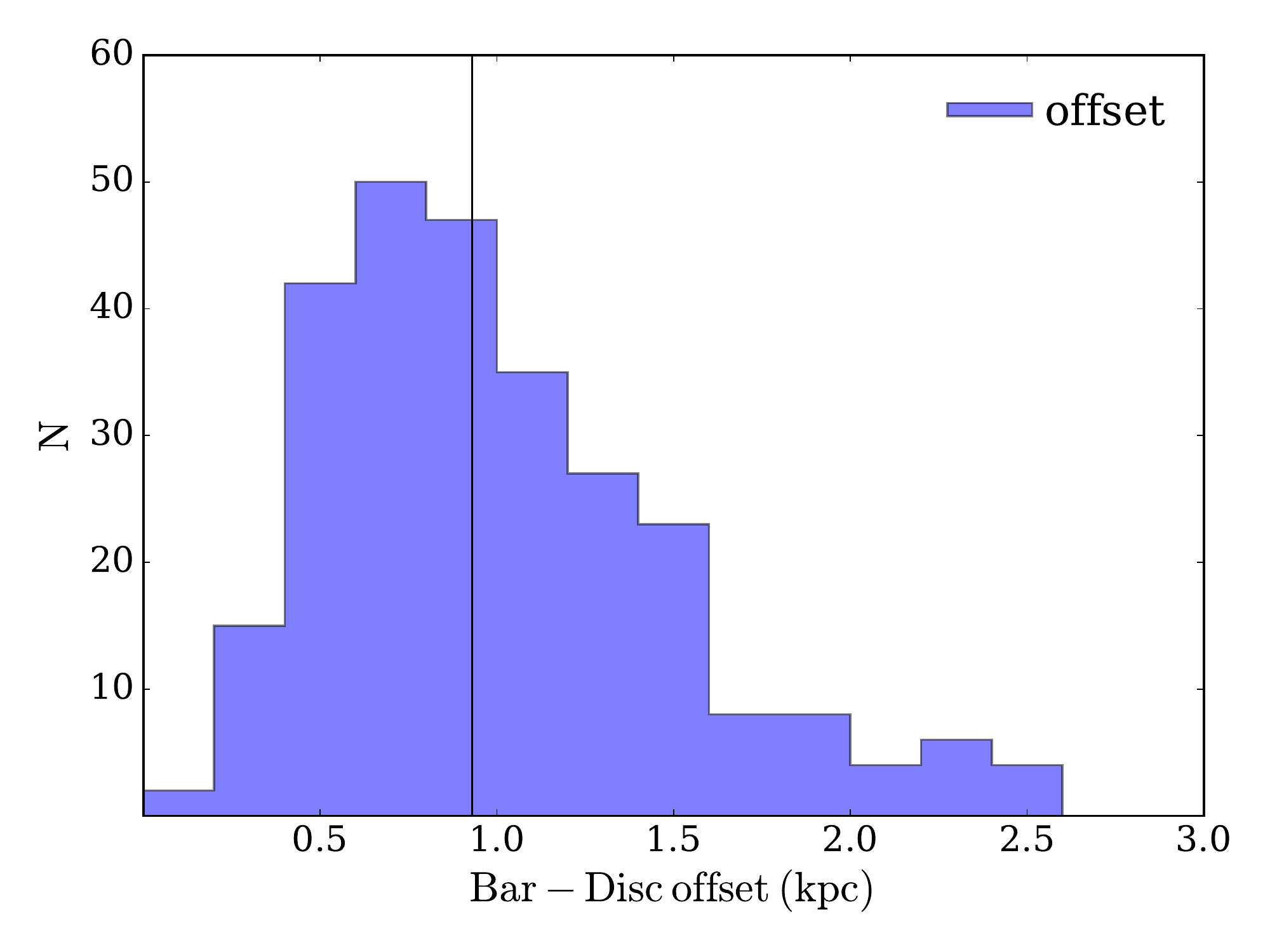}
 \caption{Distribution of the measured offsets between the photometric centres of the discs and the bars, corrected for inclination effects, in the \textsc{offset sample}. The criteria for a galaxy to have an offset bar is that the projected offset is larger than the size of the PSF. }
 \label{offset}
\end{figure}

We measured the offsets as the separation between the geometric centre of the exponential disc component and the centre of the bar component, and deprojected them using Equation \ref{eqn}. For the 271 galaxies in the \textsc{offset sample} the measured physical offsets varied between 0.2-2.5 kpc (with a median offset of 0.93 kpc and a standard deviation of 0.50 kpc), as seen in Figure \ref{offset}, a similar range to the one predicted by \citet{Pardy2016}, for different parameters of the dwarf-dwarf interaction. We find that there is only a very weak negative correlation of the measured offsets with $p_{bar}$ (Spearman $\rho=-0.16,\:p=0.01$), suggesting that our study is not biased against galaxies with the largest offsets, albeit we reiterate that we select mostly intermediate and strong bars with the selection of $p_{bar}\geq0.5$.

\subsection{Lopsidedness}

\begin{figure}
 \includegraphics[width=\columnwidth]{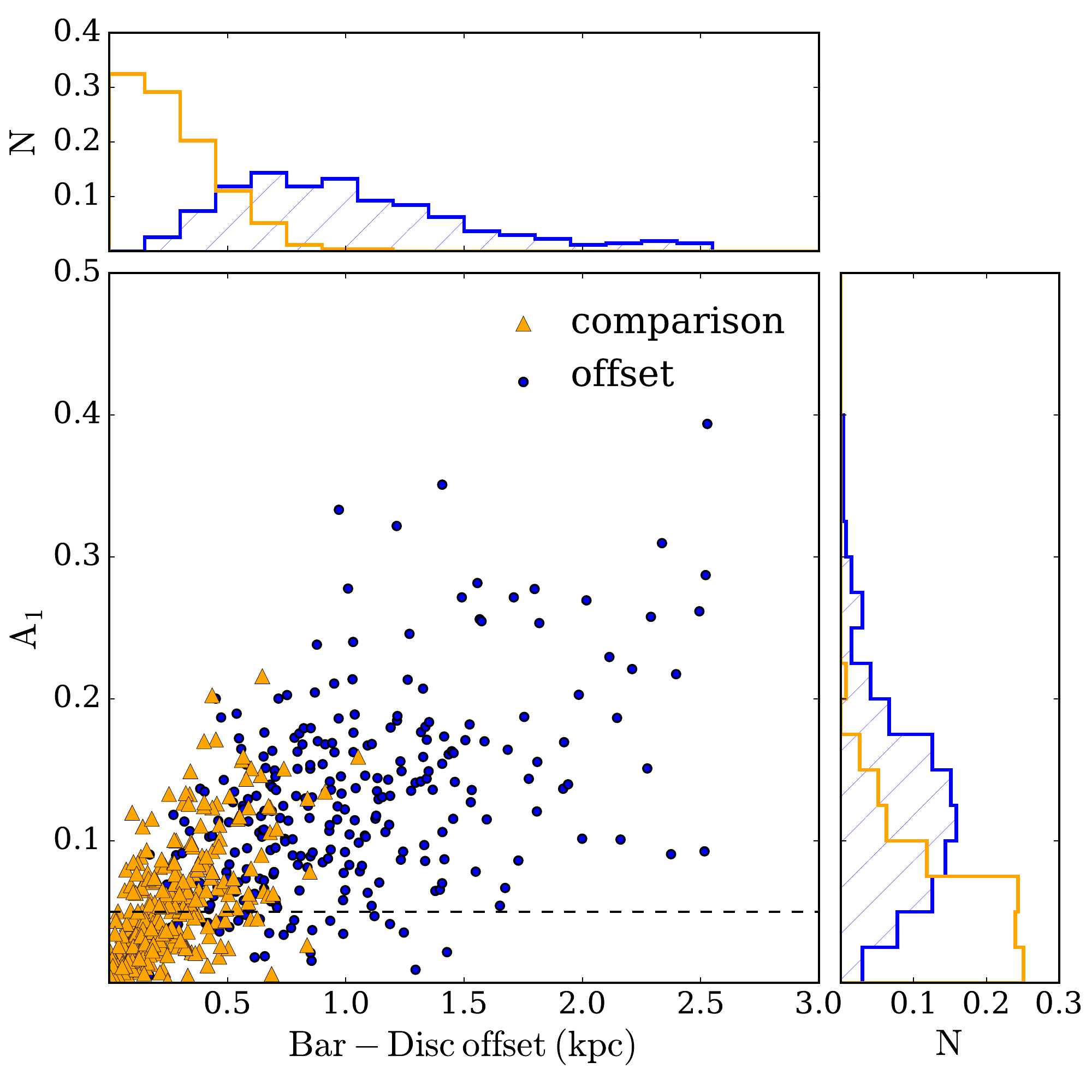}
 \caption{The Fourier $\textit{m}=1$ mode amplitude, $A_{1}$, is correlated with the offset between the disc and the bar. $A_{1}>0.05$ is an indicator of lopsidedness, shown by the dotted line in the plot. The normalized histograms show the distributions of $A_{1}$ for the \textsc{offset sample} and \textsc{comparison sample} (on the vertical) and the distribution of the deprojected offsets for the two data sets (on the horizontal).}
 \label{fourier}
\end{figure}

Using the amplitude of the first Fourier mode, $A_{1}$, as an indicator of lopsidedness (described  further in Section 3.3), we found that $A_{1}$ varies between 0 and 0.40, with a median of 0.12 in the \textsc{offset sample}. In contrast, the \textsc{comparison sample} has a median $A_{1}$ of 0.05. As expected, we find a weak, but significant correlation between the measured $A_{1}$ and disc-bar offsets (Spearman $\rho=0.4,\:p<10^{-11}$). Almost all the galaxies with off-centre bars are lopsided, with 90\% having $A_{1}>0.05$, which, according to \citet{Bournaud2005} is an indicator for lopsidedness. 63\% of the galaxies in our sample show strong lopsidedness with $A_{1}>0.10$. The strongest asymmetry in the central regions of the galactic disc is produced by the off-centre bar, and the correlation between the disc-bar offset and $A_{1}$ is seen in Figure \ref{fourier}, which matches the simulation prediction in Figure 6 of \citet{Pardy2016}. A Kolmogorov-Smirnov (K-S) test on the \textsc{offset sample} and the \textsc{comparison sample} of galaxies with centred bars gives $k=0.53$ and $p_{KS}<10^{-15}$, suggesting that galaxies with off-centre bars are more lopsided than the galaxies with centred bars.

Using a sample of 149 galaxies observed in the infra-red, \citet{Bournaud2005} has shown that the $m=1$ distortions correlate with the presence of $m=2$ spiral arms and bars, but the strong lopsidedness is not correlated with the presence of interacting companions. Furthermore, \citet{Zaritsky2013} found that nearby low surface brightness, late-type galaxies in the $\mathrm{S^{4}G}$ survey show significant lopsidedness which does not depend on a rare event, such as the accretion of a satellite. They found a similar average value of lopsidedness in local barred galaxies in $\mathrm{S^{4}G}$ survey, $\left<A_{1}\right>=0.15$, however they measured $\left<A_{1}\right>$ at the outer isophotes and not using 2D fitting. They noted that the lopsidedness is not correlated with the presence and strength of a bar as many non-barred galaxies are also lopsided, however they did not make a distinction between galaxies with off-centre bars and those with centred bars.

\subsection{Offset population properties}

Further, we desire to study the statistics of the offset population in greater detail. The survey is incomplete for fainter galaxies at higher redshifts, thus we select a \textsc{volume-limited sample}. As illustrated in Figure \ref{sample_selection}, from the \textsc{fitted-bar sample} we select only galaxies in the redshift range 0.005<$z$<0.04 and brighter than $M_{r}\le-19.22$, which is the \textit{r}-band absolute magnitude corresponding to the GZ2 completeness magnitude of 17, at a redshift $z=0.04$. We choose this redshift cut as a compromise between including fainter galaxies and having a sufficiently large sample. Because of resolution effects, it is easier to detect smaller offsets in more local galaxies, thus choosing a lower redshift limit of $z=0.04$ is justifiable. This \textsc{volume-limited sample} consists of a subset of 1,583 barred galaxies from the \textsc{fitted-bar sample}: 693 `disc dominated' galaxies (44\% of the sample) and 890 galaxies with `obvious bulges' (56\% of the sample). In this \textsc{volume-limited sample}, 8\%, or 131 galaxies are offset systems. In the following subsections we use the \textsc{volume-limited sample} and the corresponding subsample of offset systems when discussing their properties.

\subsubsection{Mass distribution}

\begin{figure}
 \includegraphics[width=\columnwidth]{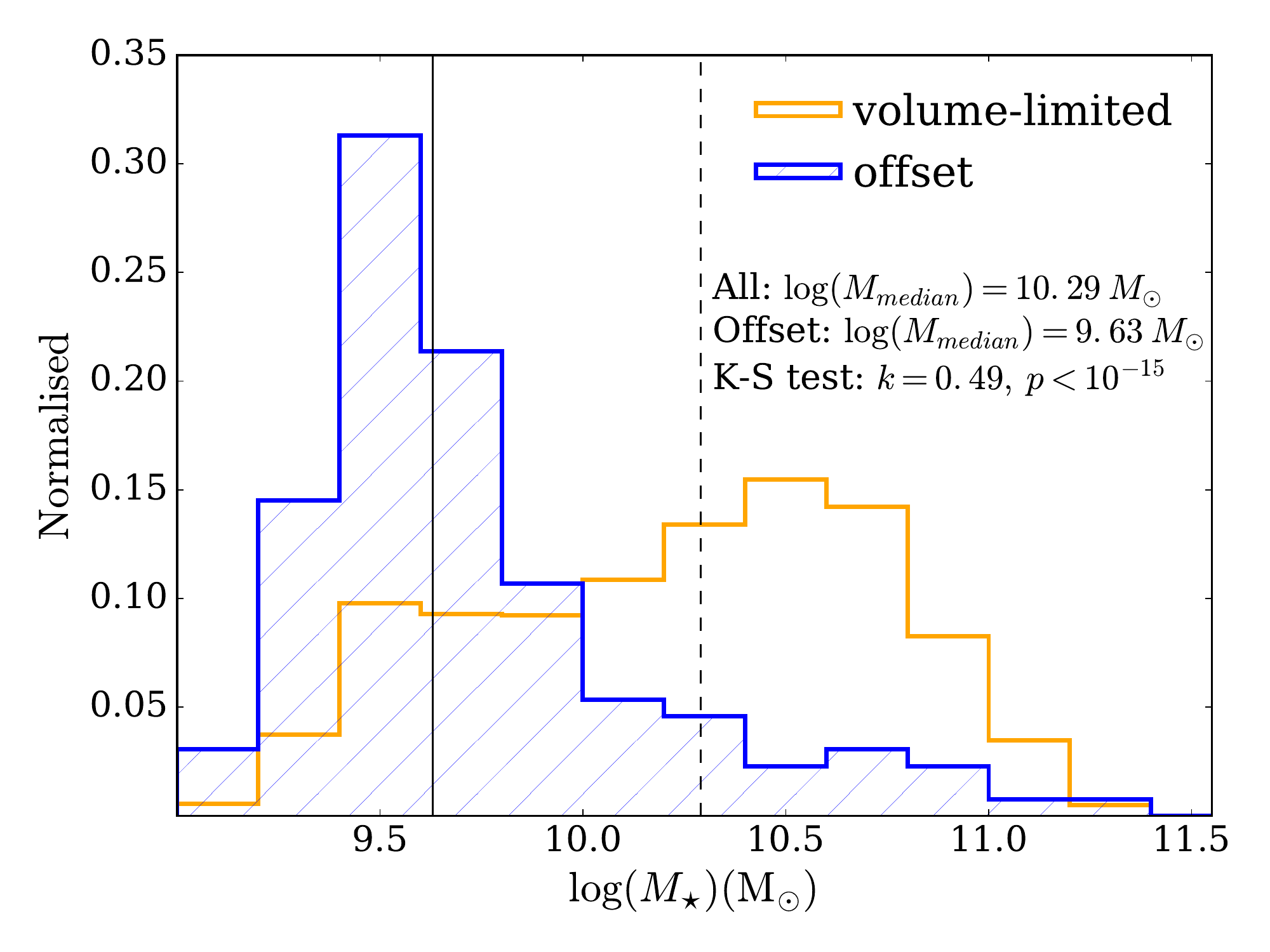}
 \caption{Normalized histograms of the mass distribution of galaxies with offset bars in the volume-limited sample (131 galaxies) and the \textsc{volume-limited sample} of barred galaxies (1,583 galaxies). The median mass of galaxies with off-centre bars is $10^{9.63} M_{\odot}$ (as shown by the vertical solid line), while the median mass of barred galaxies is $10^{10.29} M_{\odot}$ (as shown by the vertical dashed line). Only 12 galaxies with $3 \times10^{10} M_{\odot}$ are seen to have off-centre bars.}
 \label{mass}
\end{figure}

The distribution of stellar masses \citep[drawn from average values in the MPA-JHU catalogue;][]{Kauffmann2003a} for the 131 galaxies identified as having off-centre bars, as well as for the entire \textsc{volume-limited sample}, can be seen in Figure \ref{mass}. The two distributions are clearly different, the barred galaxies have a median mass of $10^{10.3} M_{\odot}$, while the galaxies with off-centre bars have a median mass of $10^{9.6} M_{\odot}$. A K-S test gives a value of $k=0.49$ and $p_{KS}<10^{-15}$; there is no evidence that the two distributions are similar. This suggests that offsets between the discs and bars are properties of lower mass barred galaxies. 

The masses of the volume-limited sample of offset systems lie between $10^{9}-10^{11} M_{\odot}$, similar to Magellanic type dwarfs, with a typical (median) mass of $4 \times 10^{9} M_{\odot}$. We find that $\sim$$20\%$ of the dwarf galaxies (with $M<10^{10} M_{\odot}$) of the \textsc{volume-limited sample} have offset bars. Furthermore, 28\% of the barred galaxies with masses between $10^{9}-10^{9.6} M_{\odot}$ have off-centre bars, suggesting that offsets are most common in barred galaxies of these masses.

We also find that only 12\% of the galaxies with offset bars have masses larger than $10^{10.3} M_{\odot}$, even though this is the median mass of the \textsc{volume-limited sample}. Furthermore, only five offset galaxies are as massive as the Milky Way (with a mass of $\sim$$10^{10.8} M_{\odot}$ \citet{Licquia2015}). This poses a challenge to simulations, as a 1:10 mass ratio interaction can be scaled up from a SMC - LMC to a LMC - Milky Way type interaction. Our observations suggest that such an interaction should not affect the relative position of the bar and disc significantly. 

Since this section concerns galaxies in a volume-limited sample, observational biases should not be responsible for the observed correlation between the offsets and lower stellar masses. It is possible that a higher fraction of even lower mass galaxies host offset bars, however more local and deeper surveys are needed to better probe the $10^{7}-10^{9} M_{\odot}$ mass range.

\subsubsection{Star formation rates}

\begin{figure}
 \includegraphics[width=\columnwidth]{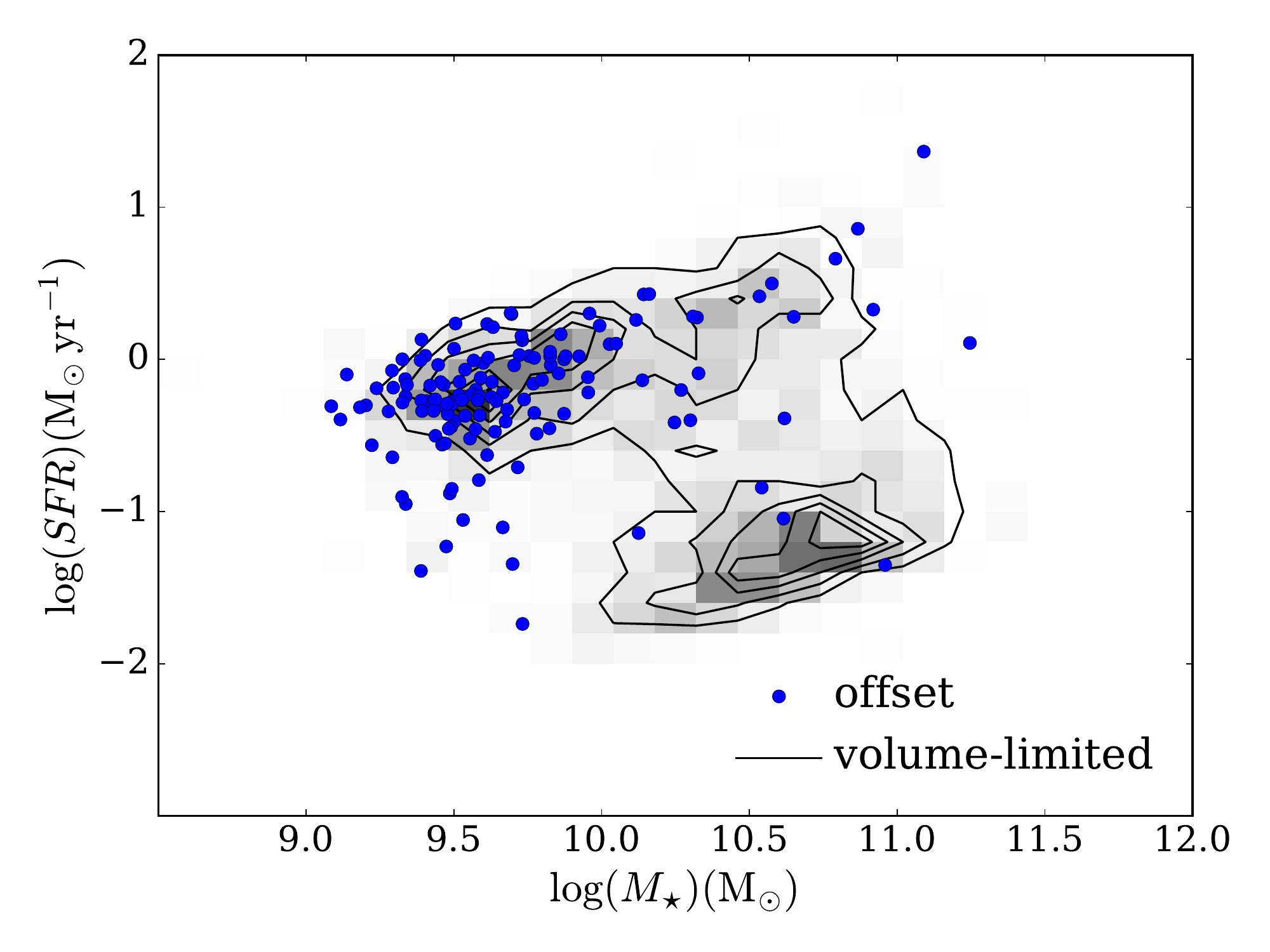}
 \caption{The location of the offset systems on a SFR-Mass plot, overlaid on the \textsc{volume-limited sample} of barred galaxies. Galaxies with offset bars are located almost entirely on the star forming main sequence.}
 \label{MS}
\end{figure}

In Figure \ref{MS} we plot the $\mathrm{SFR}$ \citep{Brinchmann2004} against the stellar mass and notice that most offset galaxies are young, blue and star-forming, being situated on the star forming main sequence, in contrast with the majority of the barred galaxies which are red in colour, as identified by \citet{Masters2011}. 21 out of 131 galaxies ($16\%$) have star formation rates below log(SFR)$=-0.5\:M_{\odot}\:\mathrm{yr}^{-1}$ and are below the main sequence, in the `Green Valley' or `Red Sequence'. Within our sample, at $M_{\star}<10^{10} M_{\odot}$, barred galaxies are typically star forming. There is no significant difference in the SFR of galaxies with offset and centred bars. We note that our \textsc{volume-limited sample} is incomplete for red (and so likely passive) galaxies at $M_{\star}\lesssim10^{10} M_{\odot}$ and, therefore, cannot rule out differences in star formation fractions at low masses.

\subsubsection{Stellar bar properties}

From the GALFITM fits it is possible to estimate the properties of light profiles of the individual components. The stellar bars in the offset-bar systems are characterized by a median ellipticity of $\epsilon=0.72 \pm 0.10$ (error bars are $1\sigma$) and they contain $0.15\pm0.09\%$ of the total light of the galaxy in the \textit{r}-band ($Bar/T$ ratio). The bars have an almost exponential light profile, of median S\'ersic index $n=0.93\pm0.70$. \citet{Kim2015} pointed out that this is indicative of a young population. They used a recent survey of 144 barred galaxies and showed that the brightness profile of the bar can be used as an indicator of its age. Bars are believed to be born out of disc material, which has an exponential profile, and in their evolution, they trap stars in the bar orbits \citep{Sellwood1993, Sellwood2014, Athanassoula2013}, flattening the light profile. 

We measure similar median values for these parameters for the \textsc{volume-limited sample}:  $\epsilon=0.68 \pm 0.13$ and $Bar/T=0.14\pm0.12$ in the \textit{r}-band. The median S\'ersic index of $n=0.67^{+1.22}_{-0.57}$ reflects the different populations of bars: bars with low S\'ersic indices in early-type galaxies and bars with close to exponential profiles in late-type barred galaxies. These suggest that the main determinant of the structure of these galaxies is the stellar mass, rather than the physical process that is causing the bar to be off-centre from the disc. 

With the fits in five different bands, it is possible to estimate the optical colours of the components, which were corrected for the dust extinction in the Milky Way, using the maps from \citet{Schlegel1998}. The discs and the bars of the galaxies in the offset-bar sample have similar blue colours, with a median $u-r $$\sim$$1.5$. Therefore, it is reasonable to assume that stellar populations of the bar are the same as those in the disc. Converting to stellar masses, we find that the typical (median) mass of the stellar bar is $\sim$$ 6 \times10^{8} M_{\odot}$, which is comparable to the mass of the bars in other Magellanic type galaxies ($3 \times10^{8} M_{\odot}$ for NGC3906 \citep{Swardt2015}, for example). 

\subsubsection{Bulges}

Only $10\%$ of the offset galaxies (14 out of 131) have `obvious bulges', while $90\%$ (117 out of 131) have `just noticeable` or `no bulges'. This is in striking contrast with the distribution of bulge types of the \textsc{volume-limited sample} of which 56\% are `obvious bulges' and 44\% are `disc dominated', suggesting that the presence of an off-centre bar is connected to the absence of a considerable bulge. Considering that half of the massive disc galaxies are barred \citep{Masters2012} and that bulges grow with the total mass of a galaxy \citep{Kauffmann2003b}, we would expect a similar fraction of offset galaxies with `obvious bulges', if stellar mass does not play an important role in the process causing the offsets. This also implies a lack of significant mergers, as even minor mergers of 1:10 mass ratio are believed to build up bulges \citep{Walker1996}. 

We test the effect of not accounting for `obvious bulges' in the fits by using the second step in the fitting procedure (disc+bar) for all the barred galaxies. In this case, the S\'ersic indices of the bars in two component fits are artificially increased compared to the three component fits (median S\'ersic index $n_{bar}=1.96$ compared to $n_{bar}=0.67$) because of the central concentration which is not accounted for \citep{Peng2010}. For the galaxies with `obvious bulges' the parameters of the bar are unrealistic in the two component fit, however, the centres of the bar and bar+bulge are approximately the same, the average offset for the whole sample of `obvious bulges' being 0.19 kpc in both cases. Therefore, using a simple bar+disc model for all the galaxies we would arrive at a similar sample of galaxies with offset bars and to the same result that the distribution of masses of offset galaxies and the volume-limited barred sample is significantly different. 

\subsection{Companions}

\begin{figure}
 \includegraphics[width=\columnwidth]{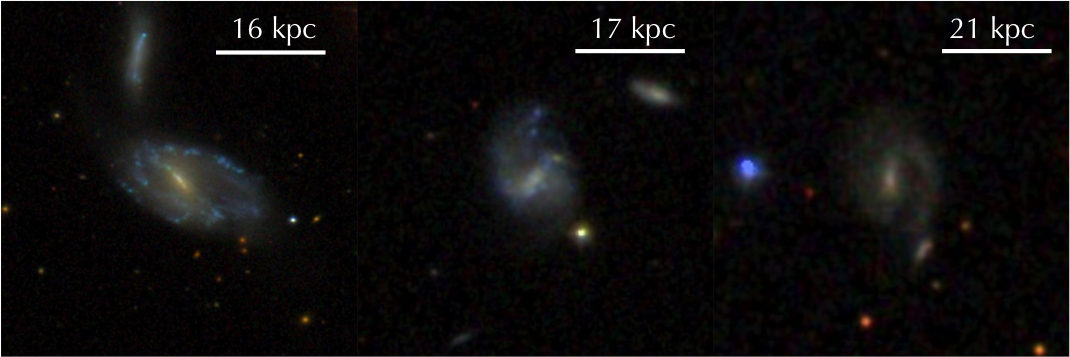}
 \caption{Examples of galaxies with offset bars that have close neighbours (<100 kpc).}
 \label{pair}
\end{figure}

In order to test the hypothesis that the offsets between discs and the bars are caused by a tidal interaction with a smaller companion, we conduct a search for such companions in SDSS, following the recently published method by \citet{Patton2016} which was also used in \citet{Barton2000}, \citet{Ellison2008} and \citet{Patton2011}. In this section we use the \textsc{fitted-bar sample}, the \textsc{offset sample} of 271 galaxies and the similar sized mass and redshift-matched \textsc{comparison sample} of galaxies with centred bars, as defined in Section 2.

We identify the closest companion for each galaxy in our samples, in SDSS, by considering as potential companions only those galaxies which have measured spectroscopic redshifts. We define a potential closest companion to be any galaxy which has $\Delta v$ within $1000 \:\mathrm{km\,s^{-1}}$ of the galaxy in question, with the smallest projected separation, $r_{p}$. Since we are interested in interactions of dwarf galaxies, we do not impose any mass ratio cut. 

We find that 642 out of the 3,357 galaxies ($\sim$$19\%$) in the \textsc{fitted-bar sample} have close companions, defined as within a projected separation of $r_{p}<100$ kpc. With a similar percentage, 17\%, 46 galaxies in the \textsc{offset sample} have a close companion, some examples of which can be seen in Figure \ref{pair}. An even higher percentage, 24\%, or 64 galaxies out of the 271 galaxies in the \textsc{comparison sample} have close companions, within $r_{p}$<100 kpc.

\begin{figure}
 \includegraphics[width=\columnwidth]{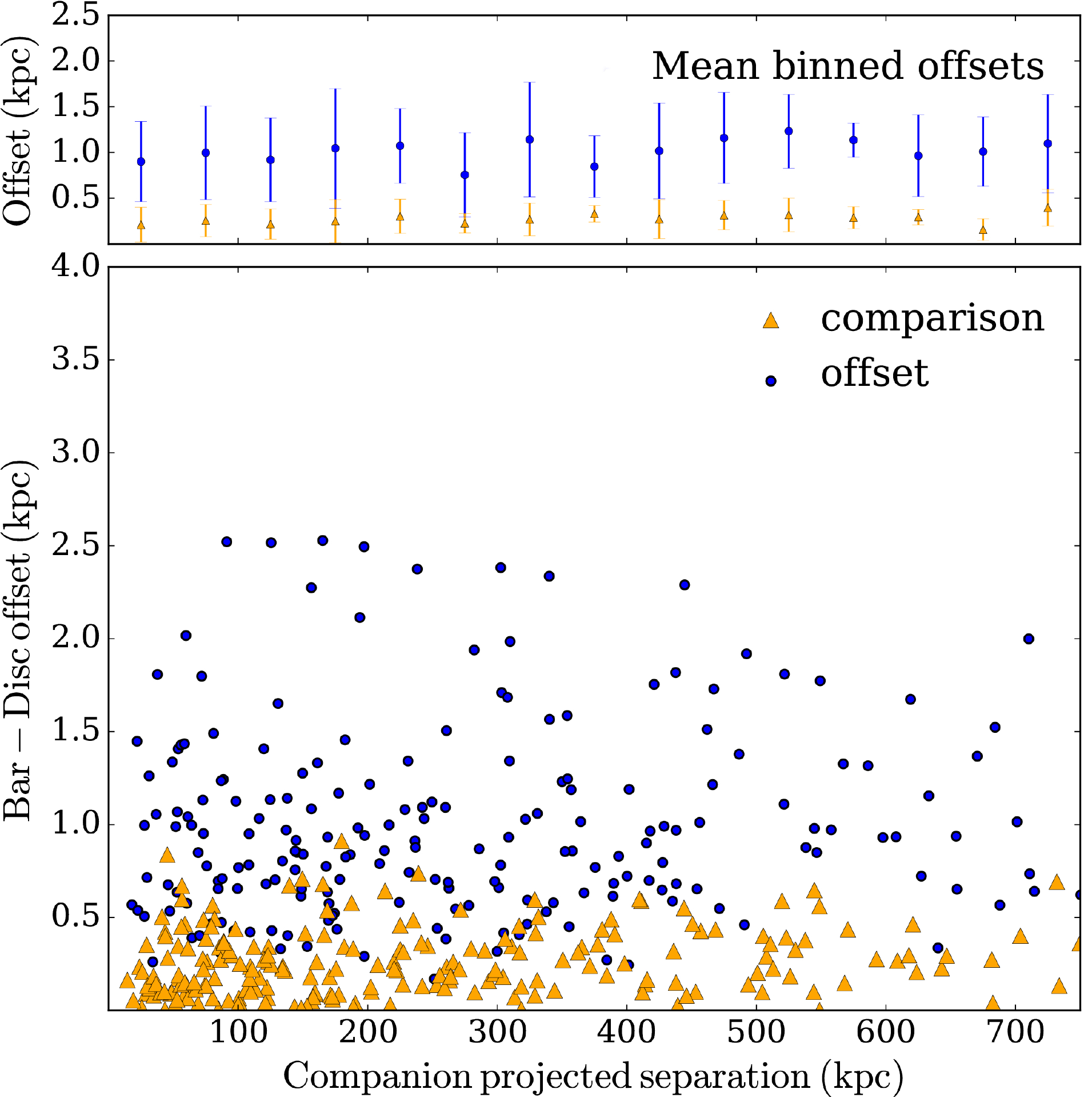}
 \caption{Bar-Disc offset versus the projected separation to the nearest neighbour with spectroscopic redshifts from SDSS. The top plot shows the same offset binned in separations of 50 kpc for the \textsc{offset sample} and the \textsc{comparison sample}. There is no clear evidence for declining offset with projected separation for the sample of galaxies with off-centre bars, the $r$-coefficient for a correlation being $r=0.17,\:p=0.01$. The mean disc-bar offset is $\sim$1 kpc across all bins for the \textsc{offset sample}. The error bars represent $1\sigma$ in each bin.}
 \label{neighbour}
\end{figure}

Simulations by \citet{Pardy2016} suggest that distortions in the disc can persist for 2 Gyr after the companion fly-by. Assuming a typical relative velocity of $375\:\mathrm{km\,s^{-1}}$ ($\sim$ LMC-SMC relative velocity), the galaxy and companion could be separated by 750 kpc at 2 Gyr after the interaction, therefore we check for companions within this projected distance. We find 199 galaxies (or 82\% of the \textsc{offset sample}) to have at least one spectroscopically confirmed companion within 750 kpc. Similarly, 86\% of the galaxies with centred bars in the \textsc{comparison sample} have at least one companion within 750 kpc.  Since the separation can be used as a proxy for the time after the interaction, we plotted the disc-bar separation versus the separation from the nearest companion in Figure \ref{neighbour} and we do not find any correlation of the offset declining with the separation, the Pearson's correlation test giving an $r$ value of 0.17.  The slight differences in close or distant companion fractions between the offset-bar sample and centred-bar comparison sample are not statistically significant. Thus, we do not find galaxies with off-centre bars to have more companions compared to similar mass barred galaxies within 750 kpc, nor closer companions within 100 kpc. There are many cases of isolated galaxies with offset bars without any apparent companion. 

It is important to note that there is high incompleteness at galaxies with small separations due to fibre collisions and deblending. The problem is especially at separations less than 55 arcsec which biases the mass and redshift distribution of close pairs \citep{Ellison2008}. This corresponds to 10-60 kpc separations in the redshift range of our sample. In the case of the brightest galaxies, the automated SDSS deblender might mistakenly identify galactic clumps as neighbouring galaxies. We inspected all the companions at the lowest angular separations, $r_{p}<30$ kpc to make sure that we are indeed detecting a companion. The nearest companion search is also incomplete because of the flux limits of the survey. The limiting magnitude for the spectroscopic survey in SDSS is $m_{r}=17.77$ \citep{Strauss2002}, where $m_{r}$ is the Galactic extinction-corrected Petrosian magnitude. The \textit{r}-band magnitudes of the galaxies in our \textsc{offset sample} range between $12.55<m_{r}<17$ and this means that there will often be low mass companion galaxies which are not detected. For example, for a galaxy of magnitude $m_{r}=16$, which is the median magnitude of our sample, we are able to spectroscopically detect a companion, if it has a mass within a factor of 5 of the primary, assuming that the observed magnitude scales with stellar mass. If we are near the faint end limit, we are strongly biased against finding less massive companions for the galaxy. We would only be able to find a 10:1 mass ratio companion for galaxies brighter than $m_{r}=15.27$. 

Based on the limiting magnitude of the SDSS spectroscopy survey, the maximum mass an unseen companion can have is $10^{8.8\pm0.4} M_{\odot}$ (median value) for the galaxies in the \textsc{offset sample}, corresponding to a median mass ratio of 5:1. Thus, it is likely that we miss companion galaxies which are 10 times less massive than the galaxies with offset bars. Deeper surveys such as SDSS Stripe 82, DECaLS \citep{Decals2015}, GAMA \citep{Driver2008} are needed to identify possible low mass companions and search for tidal features as potential evidence of minor mergers.

\section{Discussion}

Interestingly, we find a mass above which galaxies are unlikely to have offset bars, $\sim$$ 3 \times10^{10} M_{\odot}$, similar to that noted by \citet{Kauffmann2003b}, who showed that the properties of galaxies in the low redshift universe change significantly at this mass. Lower mass galaxies have younger stellar populations and are disc-dominated, while higher mass galaxies tend to be more concentrated, with higher stellar mass surface densities typical of bulges. Given that $\sim$$34\%$ of galaxies in the \textsc{volume-limited sample} have higher masses than $ \sim $$3 \times10^{10} M_{\odot}$ (as seen in Figure \ref{mass}) and only $\sim$$2\%$ of them have offset bars, it is highly unlikely that finding a similar mass threshold is a simple coincidence. We suggest that the growth of bulges, expected to happen at the same characteristic mass, stabilises the disc, preventing it from moving around the centre of mass of the galaxy. In such systems, a significant fraction of the galaxy mass is in the bulge and the bar which will produce a steeper potential well. Being highly concentrated, the inner components will reduce the self-gravity of the disc and it will prevent the disc from shifting significantly due to an interaction. This transition from a rotation supported stellar disc to a pressure-dominated spheroid can be sufficient to stabilise the disc and also cause morphological quenching \citep{Martig2009, Kaviraj2014}.

If offsets between discs and bars are truly caused by interactions with lower mass companions, another possibility for observing overwhelmingly more offsets in lower mass galaxies compared to high mass galaxies is a difference in the interaction rates. \citet{Liu2011} showed that in the SDSS survey (and similarly \citet{Robotham2012} in the GAMA survey) there is an $\sim$11\% chance for a galaxy with a similar mass to the Milky Way to have a companion at least as massive as the LMC (thus with a 10:1 mass ratio). In our volume-limited study, we find that only 2\% of the galaxies with the mass of the Milky Way (between $10^{10.5}-10^{11.1} M_{\odot}$) have offset bars, while the fraction of galaxies with masses $10^{9}-10^{9.6} M_{\odot}$ having offset bars is as much as 28\%. If an interaction is equally likely to cause an offset bar, regardless of the mass of the main galaxy, it is very improbable that the interaction rate for low mass galaxies is so much higher.

Even though we do not find a correlation between the galaxies with off-centre bars and the nearest companions, tidal interactions between the galaxy and a small companion, as suggested by \citet{Pardy2016}, cannot be ruled out. The incompleteness due to the flux limit of SDSS and fiber collisions at the smallest separation make the closest spectroscopic companion hard to identify. Future spectroscopic observations of potential candidate companions should be able to help identify physical companions. Another possible explanation for the missing companions are high velocity dwarfs on eccentric orbits that are now too far away to appear associated with the primary galaxy, on the long timescales in which the offset is restored. Further simulations of dwarf-dwarf interactions that better explore the parameter space (mass ratios, relative velocities, impact parameters, collision angles) are needed to quantify the disc-bar offsets and constrain how long the offset lasts in different galaxy interactions.

Despite not being able to identify all the physical companions, the large number of isolated galaxies with off-centre bars in our sample and other studies \citep{Feitzinger1980, Wilcots2004} is puzzling. We should consider a different explanation for the offsets seen in some galaxies. One suggested origin is the interaction with `dark' satellites, with no or very few stars \citep{Bekki2009}. Another plausible explanation is the asymmetry of the dark matter halo \citep{levine1998} or the misalignment between the stars and the dark matter halo. The dark matter halo is far more massive and more extended than the galactic disc, thus it is more susceptible to distortions. If galaxy interactions are common, we should expect them to primarily have an effect on the dark matter haloes. Lopsided haloes may also form via the accretion of dark matter following the cosmological perturbations. The dynamics of stars in a galactic disc as a response to a perturbed halo potential has been studied by \citet{Jog1997} and \citet{Jog1999} and has been shown to lead to lopsided discs, such as the discs of M101 and NGC1637 \citep{Sandage1961}. Since we find a correlation between the off-centre bars and the galaxies being lopsided, the asymmetries in the dark matter halo could also lead to the observed offsets and this might explain the missing companions. With future observations of the kinematics of these galaxies with resolved integral field spectroscopy, such as the MaNGA survey \citep{Manga2016}, we will be able to directly determine the dynamical centre of the galaxies and this could shed light on the mass distribution of the galactic halo. 

\section{Conclusion}

We identified a sample of 271 barred galaxies in SDSS with an offset bar from the photometric centre of the disc and selected a volume-limited subsample to study the properties of these systems. Our study used morphological classifications from the Galaxy Zoo project and 2D photometric decompositions and is the first systematic search for such systems. The vast majority of these galaxies have similar properties to the Large Magellanic Cloud: similar masses, optical colours and measured bar offsets. These galaxies are highly asymmetric, and the offsets between the disc and the bar are an explanation of their lopsidedness. Our observations show that there is a mass of $3 \times10^{10} M_{\odot}$ above which the galactic discs are stable against disc-bar offsets, only $2\%$ of the barred galaxies above this mass showing offsets. This mass transition should be explained by future simulations. It is believed that these offsets trace minor interactions, however, we do not find statistically significant evidence of a correlation with the nearest companions, even though the measured physical offsets match the predicted values from simulations of tidal interactions. This could be due to the incompleteness of the SDSS spectroscopic survey at the faint flux limit and observations of possible companion candidates should be done in order to confirm their spectroscopic redshifts. Many isolated galaxies show evidence of an offset bar, which cannot be attributed to a dwarf-dwarf interaction. Other possible explanations for the offset should also be considered, such as an interaction with a dark matter subhalo or an asymmetry in the dark matter distribution in the halo.

\section*{Acknowledgements}
The authors would like to thank the anonymous referee for their helpful and thoughtful comments. The authors would also like to thank C. Jog for the insightful discussion on the alternative scenarios that can give rise to off-centre bars.

SJK acknowledges funding from the Science and Technology Facilities Council (STFC) Grant Code ST/MJ0371X/1. KS acknowledges support from Swiss National Science Foundation Grants PP00P2\_138979 and PP00P2\_166159. RJS acknowledges funding from the STFC Grant Code ST/K502236/1. Support for this work was provided by the National Aeronautics and Space Administration through Einstein Postdoctoral Fellowship Award Number PF5-160143 issued by the Chandra X-ray Observatory Center, which is operated by the Smithsonian Astrophysical Observatory for and on behalf of the National Aeronautics Space Administration under contract NAS8-03060. 

The development of Galaxy Zoo was supported in part by the Alfred P. Sloan Foundation and by The Leverhulme Trust. 

Funding for the SDSS and SDSS-II has been provided by the Alfred P. Sloan Foundation, the Participating Institutions, the National Science Foundation, the US Department of Energy, the National Aeronautics and Space Administration, the Japanese Monbukagakusho, the Max Planck Society, and the Higher Education Funding Council for England. The SDSS website is \url{http://www.sdss.org/}. The SDSS is managed by the Astrophysical Research Consortium for the Participating Institutions. The Participating Institutions are the American Museum of Natural History, Astrophysical Institute Potsdam, University of Basel, University of Cambridge, Case Western Reserve University, University of Chicago, Drexel University, Fermilab, the Institute for Advanced Study, the Japan Participation Group, Johns Hopkins University, the Joint Institute for Nuclear Astrophysics, the Kavli Institute for Particle Astrophysics and Cosmology, the Korean Scientist Group, the Chinese Academy of Sciences (LAMOST), Los Alamos National Laboratory, the Max-PlanckInstitute for Astronomy (MPIA), the Max-Planck-Institute for Astrophysics (MPA), New Mexico State University, Ohio State University, University of Pittsburgh, University of Portsmouth, Princeton University, the United States Naval Observatory and the University of Washington.

This research made use of NASA's Astrophysics Data System Bibliographic Services. This work made extensive use of \textit{astroPy}\footnote{\url{http://www.astropy.org/}}, a community-developed core Python package for Astronomy \citep{Astropy} and of the Tool for Operations on Catalogues And Tables, \citep[TOPCAT\footnote{\url{http://www.star.bris.ac.uk/~mbt/}};][]{Topcat}.




\bibliographystyle{mnras}
\bibliography{references} 

\bsp	
\label{lastpage}
\end{document}